# Building on the Case Teaching Method to Generate Learning Games Relevant to Numerous Educational Fields

Iza Marfisi-Schottman
SICS and LIP6, Université Pierre et Marie Curie
Stockholm, Sweden
iza@sics.se

Jean-Marc Labat, Thibault Carron
LIP6, Université Pierre et Marie Curie
Paris, France
jean-marc.labat@lip6.fr, thibault.carron@lip6.fr

*Abstract*—University teachers often feel the need to try innovative learning technologies such as Learning Games to motivate the new generation of students. However, the typically limited resources of universities coupled with the high cost of designing and developing Learning Games result in it rarely being feasible to meet this need. To address this challenging problem, we have designed a framework that allows teachers to create their own Learning Games with very little or no help from developers and graphic designers. This framework, tested and validated by several university teachers, is suited to a wide variety of educational fields because it generates Learning Games based on the widely-used case teaching method.

*Keywords— Serious Games; Learning Games; genericity; design; authoring tool, case method; case study; adaptability*

## I. THE GENERIC SERIOUS GAME PROJECT

An increasing number of university teachers feel the need to integrate Learning Games (LGs)into their courses to motivate the new generation of students. LGs are computer applications that use game mechanisms such as competition, rewards or simply curiosity to captivate the learner's attention and ease their learning process [1]. Numerous new LGs have been created for higher education these past years: the Laboratorium of Epidemiology [2] uses a LG to trainmedical students to find solutions to combat the propagation of nosocomial diseases in hospitals, Prog&Play [3] is a LG designed to teach programming skills and many other LGs have recently been created to teachskills in various domains such as mechanisms of financial transactions, machinery management, company productivity, data security…

However, these LGs often represent a big investment that is far from been accessible to all universities, especially given that they often target a very precise topic, making it very difficult to have a positive return on investment. The few authoring tools that exist [4,5] are either not available to teachers, or are too complex to use or require development skills. In addition, these LGs have each been designed and created for a specific teacher,and they are then very rarely used by other teachers. This is because colleagues do not have the means to adapt an existing LG to their personal way of teaching in order to feel comfortable using it [6].

Toaddress these challenging problems, the French government has decided to support the Generic Serious Game project that brings together research centers, private IT companies and six thematic digital universities that teach science, medical practice, social science, economy and law.

The main goal of this project is to design and develop an authoring framework that will allow university teachers to create their own LGs with very little or no help from developers and graphic designers [7]. Although such a framework is a good solution to lower the cost of LGs, it raises the challenge of designing a framework that can generate LGs that are relevantto thelarge variety of educational domains taught in universities.

In this article, we present the current findings of this project. We start by presenting our analysis of the case teaching method and show that it can be applied to a large number of educational fields. Then we put forward our main proposal: **GenCSG** (Generic Case Study Game), a LG authoring framework based on the case method that also supports adaptation to various teaching situations. Finally, we will review the validation process of our proposal that was used by several university teachers involved in the project and put forward a few of our research perspectives.

## II. CASE METHOD : A GENERIC TEACHING TECHNIQUE

Although it seems impossible to find a LG scenario that would suit all of the educational fields taught at university, we have identified one teaching technique that is used in almost all universities and all fields of education: the **case method**. This teaching technique consists in presenting the learner with a problem (the case), inspired by a real situation, and putting him in the position of the decision maker. This simple and effective way of learning, that was first established as a teaching method with a formal structure at Harvard Business School [8] in the 1970s, is now used in many domains[9]and is relevant to all types of pedagogical theories [10]. Let us present the various steps that we have identified for case studies and illustrate them with two very different domains: medical emergency and law .

1. Acknowledge the problem
Take note of the patient's problem and the initial information (age, family history, symptoms, allergies…)
Take note of the client's situation and problem

2. Until you consider that you have sufficient information to make a decision,
   - Collect more information
Stabilize the state of the patient if necessary, interrogate him and perform medical exams
Question the client and the actors involved
   - If necessary, analyze the information collected
Analyze the results of the blood tests, listen to the heart beat to detect arrhythmia…

⚖ Extract all the relevant information from documents and interviews and reformulate it using the correct law terminology
- Referring to the new information and analysis, make inferences about the problem and add, delete, strengthen or weaken the possible solutions
  ✚ Determine the probability of each pathology
  ⚖ Analyse the rules of law that can be applied to the client's situation and determine the best solution

3. As long as more than one solution is still possible
   - Chose one solution (the most likely or the easiest to validate or invalidate)
     ✚ Choose the pathology that seems the most plausible or that is the easiest to validate or invalidate with medical exams
     ⚖ Choose the solution that seems the best for the client
   - Do further investigations to validate or invalidate this solution
     ✚ Ask for complementary medical exams (blood analysis, urine test, MRI…)
     ⚖ Investigate further to find facts that will confirm or infirm the chosen solution

4. When there is only one solution left or when one solution has been validated
   - Provide the final answer for the case study
     ✚ Diagnose the patient, prescribe treatment or send the patient to a specialized medical ward
     ⚖ Formulate the client's problem in terms of law, inform the client of the possible outcomes and gather all the relevant facts for the trial

This generic case study protocol is the base that we have chosen to build our framework on. In the next section, we will present our adaptable game design that enhances this pedagogical structure into a LG.

### III. ADAPTABLE GAME DESIGN FOR CASE STUDIES

In this section we will describe GenCSG (Generic Case Study Game) our LG authoring framework created to help teachers design case study LGs that are adaptable to their various teaching situations. First we will describe the game design we have chosen for the case study and then the authoring tools that are provided to the teachers to create and adapt their LGs.

The choices concerning the game design and the authoring tools of GenCSG were made in close collaboration with the university teachers involved in the project and reflect their specific needs. We were also strongly inspired by the experience acquired during two projects lead by our research laboratory. The first project, Play&Cure, is a LG designed to train students to perform case studies on patients with edema. This LG was designed in close collaboration with medical practitioners and has been successfully used in teaching environments. The second project, Ludiville [11], is a LG based on case studies designed to train bank employees to analyze clients' banking situations and select the relevant money loans with the suitable interest rates.

*A. Game design*

As shown in Figure 1, the game interface is composed of four distinct canvases. The canvas in the center is the main working space. It has three tabs that allow the learner to freely navigate through all the steps of the case study protocol presented in part II.

The **Problem** tab presents the case study, the problem that has to be solved and the initial textual and media information. This tab allows the learner to acknowledge the problem (step 1 of a case study). Figure 1 shows an example of the problem tab for a medical emergency case study. It contains textual information about the patient along with a photo and a video of his arrival at the hospital.

Figure 1. Presentation of the case study problem

The **Actions** tab allows the learner to collect more information about the case by doing various actions (step 2 and 3 of a case study). The actions are represented by cards grouped in categories. For the medical emergency case study (Figure 2, top), the actions are in five categories: stabilizing actions, questions, physical examinations, generic and specific explorations such as MRI scans and other medical tests. Once the learner has chosen an action category, he is presented with each available action and its "price" for each of the **scoring elements** (visible in the top right canvas). For the medical emergency case study (Figure 2, middle), the actions only have a cost in "time" because it is the only scoring item chosen by the teacher. As we will see in the next part of this article, the scoring items and mechanisms can be customized by the teachers so that they serve their pedagogical goals. When the learner chooses to do an action (by clicking on the corresponding card), the game updates the scoring items and shows the result of the action. This result can be a straightforward text such as "The dipstick test indicates no anomalies" or a more complicated result such as a video, an x-ray or a sheet with the results of a blood analysis that the learner has to analyze. For example, the result of the action "listen to heart beat", in the medical emergency case study (Figure 2, bottom), is a sound file. The teacher also added an analysis question in which the learner has to indicate if the heartbeat is normal, irregular or has a murmur. Once again, the nature and the complexity of both the result and the analysis question are totally adaptable to the teacher's needs and will therefore vary according to the resources available and the pedagogical goals. All the information obtained by the learners during their investigation is automatically collected in the **Directory** canvas on the right of the interface.

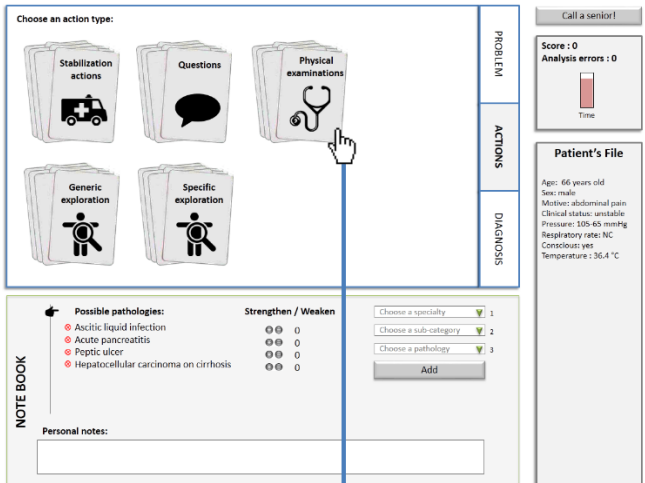

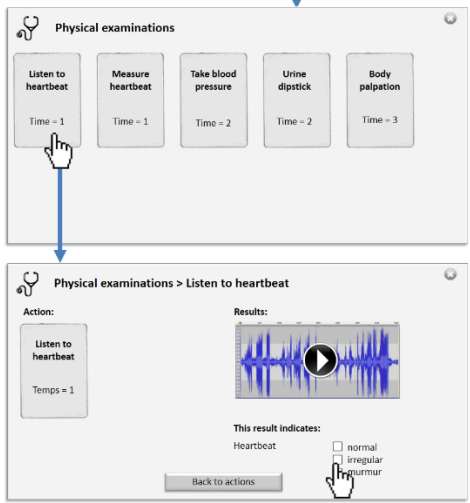

Figure 2. Player actions to invastigate the case

The canvas on the bottom of the interface is the learners' personal **Notebook** on which they can make a list of the various solutions that they are considering for the case study. Based on the information that the learners get during the investigation, they can add marks to strengthen or weaken the validity of each solution.

When the learners have collected enough information, they can enter the final answer to the problem on the **Diagnosis** tab (step 4 of a case study). For the medical emergency case study (Figure 3), the learner has to indicate the pathology that the patient is suffering from, the medical ward to which the patient should be sent, the prescription and any extra information on the pre-emergency care. When the learner clicks on the validate button, these answers are compared to the correct ones. The game also indicates the actions that were not useful or not done in the correct order and the mistakes made on the analysis questions.

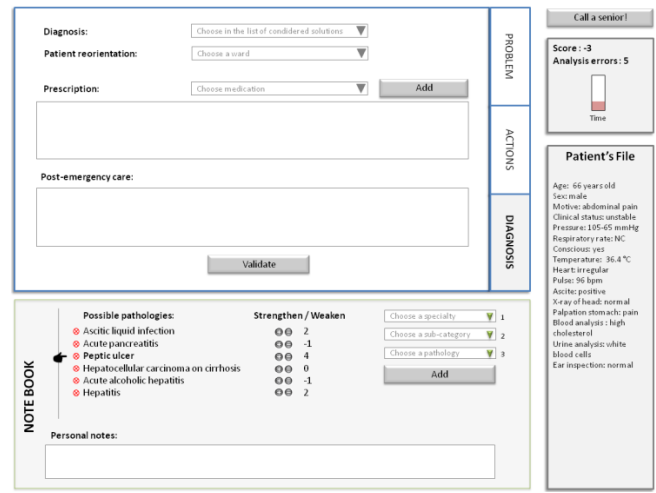

Figure 3. Final diagnosis of the case study

The structure of this game design makes it suitable for all types of case studies. It can, for example, adapt perfectly to the case study of a plane engine that broke down, the case study of a murder or of a polluted city that wants to set up ecological infrastructures. All of these case studies have a problem that can be documented with text and media, various actions that the learners can do to investigate and obtain more information and a list of possible solutions to choose from.

As you may have noticed, the game mechanisms integrated in this game design are quite minimal. This is primarily due to the request of the teachers that where involved in this project. Indeed, they wanted a simple tool that they would feel at ease manipulating and they certainly didn't want to feel imprisoned by any game format they do not fully master. The other reason that explains the minimal game mechanics, is that it is impossible to find a game scenario that would suit all the domains and contexts that university teachers deal with. His is why we designed this simple game design but, in order to help the teachers enhance their course with additional game mechanisms when they feel up to it, we also offer a set of additional gaming options that we will present in the next part of this article.

### B. Case study authoring tool

As we have seen in the first part of this article, one of the requirements formulated by university teachers is the possibility to create LGs in a very simple way and without any programming skills. To meet this requirement, we first discussed the matter with several teachers involved in the project, and then came up with a simple authoring tool based on a formatted spreadsheet that can be used by the teachers or domain experts to create new case studies. We noted that the teachers preferred to use a simple spreadsheet rather than an online authoring environment because they were used to working with such a tool and felt comfortable copying, pasting, sharing and even using the built-in validation forms.

Our authoring spreadsheet contains all the necessary information to automatically generate an operational LG. It is composed of predefined cells in which the authors have to enter the initial information about the case, the list of possible solutions and indicate the ones that are correct. The author can also create an action by adding a new line in the action section of the spreadsheet. This line has to contain the name that will appear on the card, the initial state of the card (visible, invisible or disabled), a text or a link to one or more media items in a specific folder (image, video, sound or link to a web page) that will be shown to the learner when the action is selected and a multiple-choice analysis question if necessary. The author also indicates if the action should be done or not and the impact that it will have on the learner's grade and on the various scoring items of the game. The author can also add functions(computer program methods) that will be triggered when the action is chosen by the learner. These functions can interact with the elements of the LG by hiding or enabling action cards for example or changing the scoring points of the player. In addition, the authors can change the text of almost all the labels on the LG interface so that they are formulated in an appropriate manner for the domain in question. Finally, each spreadsheet also contains a section with metadata in which the author can enter the name of the case, date of creation, author, level of difficulty, field of educational and a few sentences providing a description and suggestions as how to use the case study in class.

*C. Teacher platform for creating Learning Game sessions*

In addition to the spreadsheet that is used to create the content of the case studies, we have designed an online platform with which the teachers can create their LG session with one or several relevant case studies found in the case database by using the metadata tags. This platform also offers a certain number of pedagogical and gaming options so that the teachers can adapt the LG to their specific teaching situation:

- Let the learner choose from a list of case studies, ask the LG to randomly select the case studies from a list or ask the LG to "play" a list of case studies in a given order.
- Show the right answer to the learnersimmediately after they make a mistake or at the end of the case study.
- Show the impact on the scoring items immediately after the actions or at the end of the case study.
- Publish the scores online or not, by group or by student. Note that even if the teacher has set the game to publish all the scores by default, each student can use the options setting of the game to choose to hide their own score.

IV. EVALUATION

To validate the fact that GenCSG can easily be used to create case study LGs in a wide range of domains, we asked several teachers to each create two case studies. The experimentations usually took place in the teacher's office and started with a general presentation of the framework after which we helped them create a case study by filling in the spreadsheet. After we had collected enough content for the first case study, we created a mockup of the LG so that the teachers could see what the final result would look like. This allowed the teachers to identify certain errors and correct the elements on the spreadsheet but also to imagine use case scenarios. We then asked the teachers to work on creating the second LG on their own which was a good way to identify their difficulties and improve ourauthoring textbookguide.

All of the teachers were very enthusiastic about the project and they understood the principles of GenCSG very quickly. When they saw the LG examples, two of the teachers even had the idea of transcribing old paper format case studies they were using in class which we had no problem doing. As we will see in the next section, the teachers also thought ofseveral innovative ways of using their new LGs in class. Table I shows the various texts and items chosen by the teachers for their case study games.

The**medical practitioners**  (Paris V University) have been involved in the project from the beginning and helped us design the authoring spreadsheet. They want touse GenCSG as a way to help teachers plan the practical coursesessions in advance with a selection of five or sixshort cases for a given specialty and level.The **law** teacher (Lyon III University) has a very different approach. He would like to use case studies as a main example that can be used throughout a course and discussed with the teacher during the practical sessions.He has been working on a divorce case study for the family law course. In order to increase the flexibility of GenCSG that the teacher found to

TABLE I. TEXT AND ITEMS CHOSEN BY THE TEACHERS FOR THEIR CASE STUDY GAMES IN VARIOUS DOMAINS

| Texte and items in the LG | Problem label | Solutions label | Help label | Repository label | Available action cards | Scoring elements |
|---|---|---|---|---|---|---|
| Medical emergency | Initial state of admission | Pathologies | Call a senior | Patient file | Stabilization actions, questions, physical examinations and further explorations | Accuracy and time of the medical care |
| General practitioner | Patient's requests | Pathologies | Help | Patient file | Dialogue, physical examinations and further analysis | Accuracy, time and cost of the medical care and treatment |
| Law | Problem | Legal qualifications | Use a joker | Client file | Rules of law and facts | Accuracy of the reasonning and choice of legal qualifications |
| Mechanics | Machine failure | Investigation leads | A helping hand? | Technical report | View documentation, dissasemble and analyze | Accuracy of the diagnosis and cost of the investigation |

"determinist", we added the possibility for the students to enter their own answers in addition to the ones listed by the case creator. We also worked on the possibility of adding a chat tool to promote discussions among the students and among different groups. The **mechanics** teacher (Douai engineering school) worked on a case study in which the students have to understand a complex piece of machinery and identify the causes of a failure. This case study combines several concepts viewed in class and could be used as a midterm exam. Our partners specialized in **medical emergency** training (instructors at the emergency medical service in Marseille and Lavoisier) have been very enthusiastic about the project and the interns have started creating a dozen cases based on real patients. The team has also been working on the addition of an introductory 3D environment of an emergency ward as a way to immerse the learners in the game before they start the case study. They are designing 3D scenes as well that could be integrated into the case study for the physical palpation actions, for example (Figure 4).In addition, asLudiville, a LG for training **bank** employees, was an inspiring models for GenCSG, we were able to transpose its case studies into our system very easily.

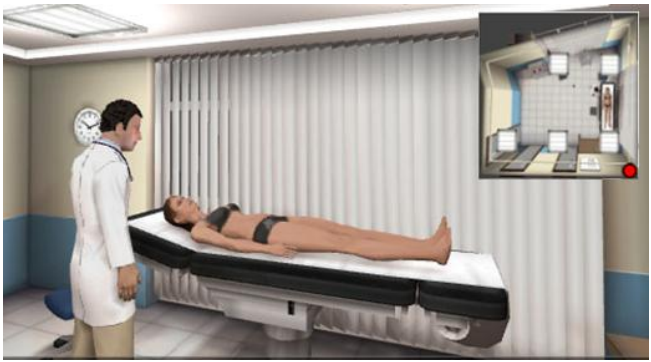

Figure 4. 3D environnement designed by Strass/Kokopelli that can be integrated into the case study or used as a shell around the case study

## V. CONCLUSION AND PERSPECTIVES

In this article we have presented GenCSG(Generic Case Study Game), a LG authoring framework that enables teachers with no programming skills tocreate case study LGs with very simple graphicsand adapt them to their specific teaching situations. In the event that the teachers have the resources to work with graphic designers and developers, our framework gives them the possibility of adding functions,that are triggered during the LG, and that can interact with external computer program objects such as a machine simulation or a 3D simulation of a human body. This system offers countless possibilities of enhancing and extending the game design easily.

To validate the fact that GenCSG can easily be used to create case study LGs in a large variety of domains, we asked an emergency doctor, a medical general practitioner, a law teacher and an engineering teacher to each create two case studies with it.

The LG generator and the authoring platform are currently being finalized and will be freely available for the85 French universities and 32 schools for higher education that have participated in the Generic Serious Game project. We are also currently working on extra game motivation aspects that could be added to the LG design such as a battle mode which would allow two students to dual over a case study. In addition, we are looking into the possibility of including a monitoring tool for the teachers that they could use during the LG session to view the student's progression and the actions in real time. This tool could help the teachers to rapidly identify the students that are having difficulties but would also allow them to trigger predefined events, an approach that has been experimented with [6], in order to add extra obstacles that challenge the faster students.

ACKNOWLEDGMENTS

The research published in this article was carried out for the Generic Serious-Game project (www.generic-sg.fr) and we thank all the actors that were involved: 6 thematic digital universities (UVED, UNISCIEL, UNF3S, AUNEGE, UNIT, UNJF), Lavoisier, Strass productions and Kokopelli.

Part of this work was carried out during the tenure of an ERCIM "Alain Bensoussan" Fellowship Programme. The research leading to these results has received funding from the European Union Seventh Framwork Programme (FP7/2007-2013) under grant agreement n$^o$246016.